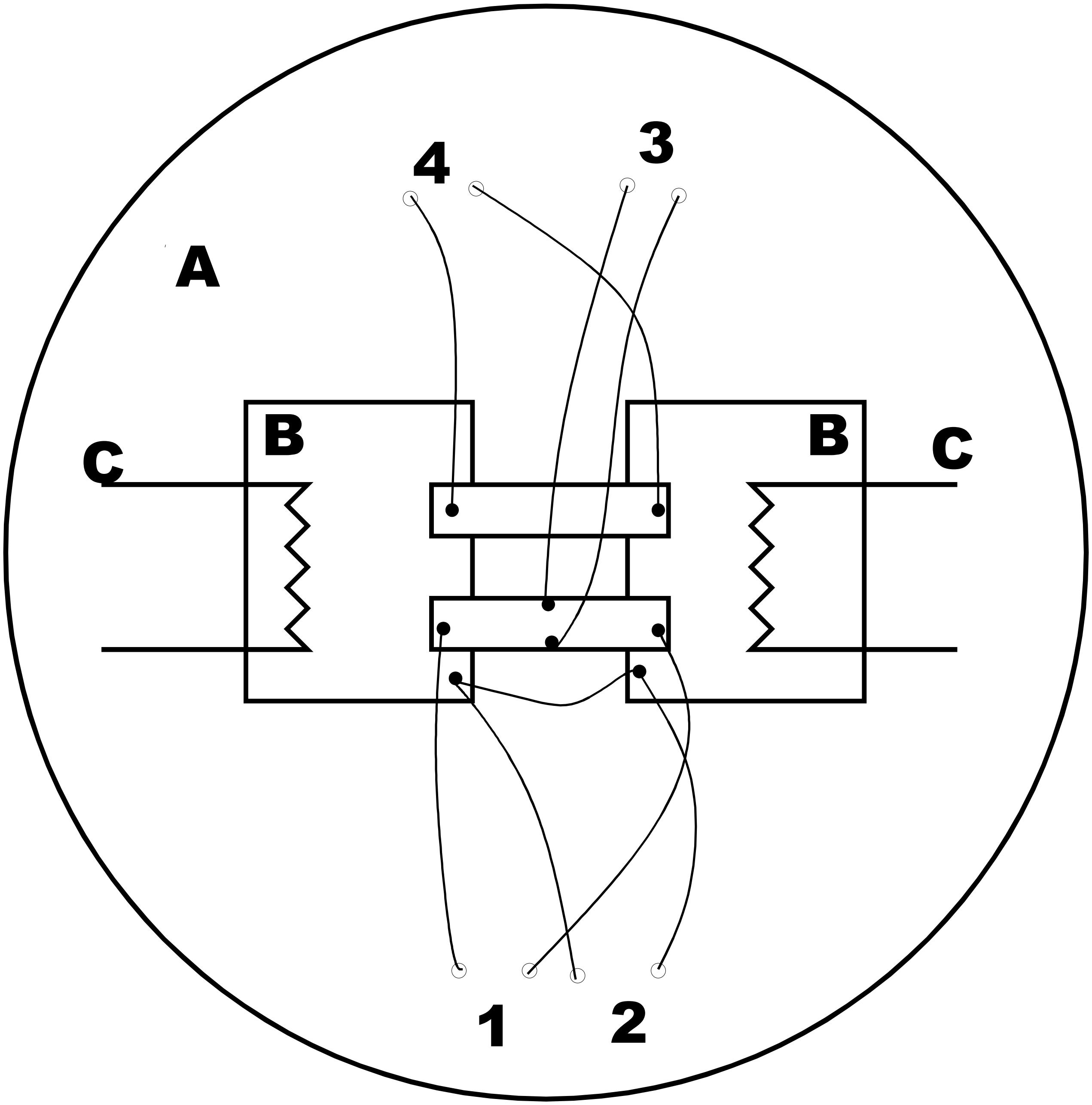

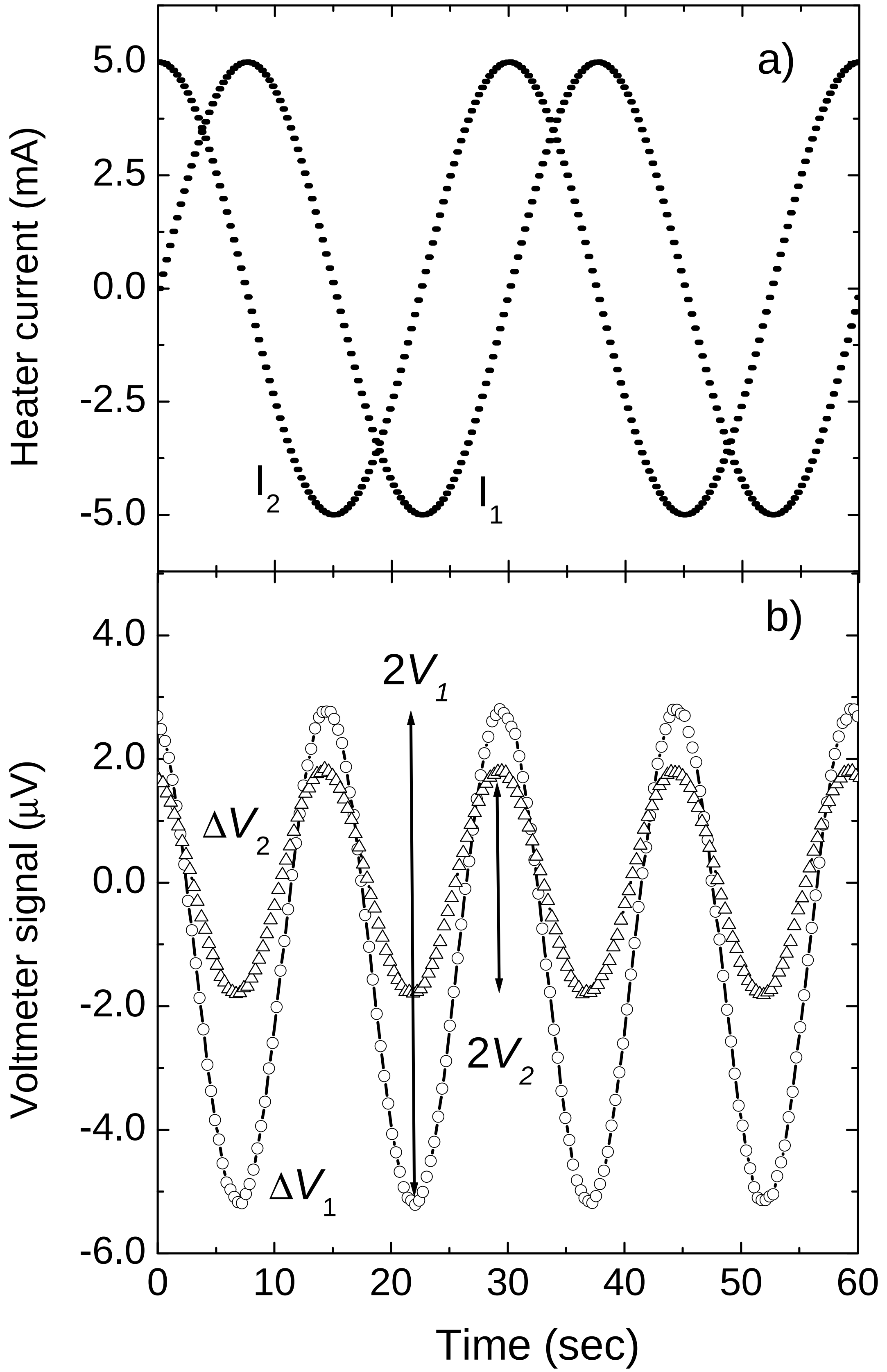

Choi et al., Fig. 2 a&b

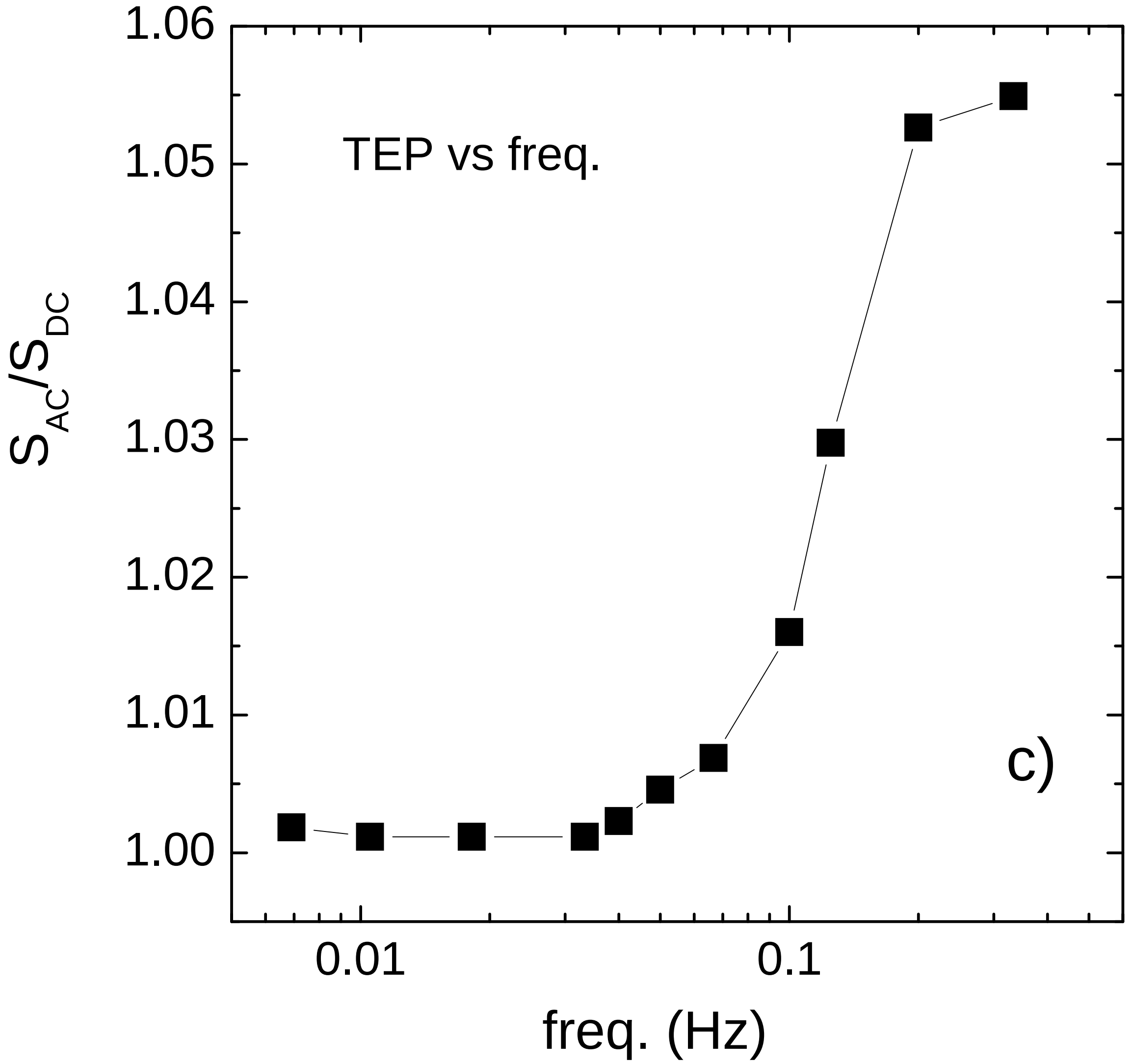

Choi et al., Fig. 2c

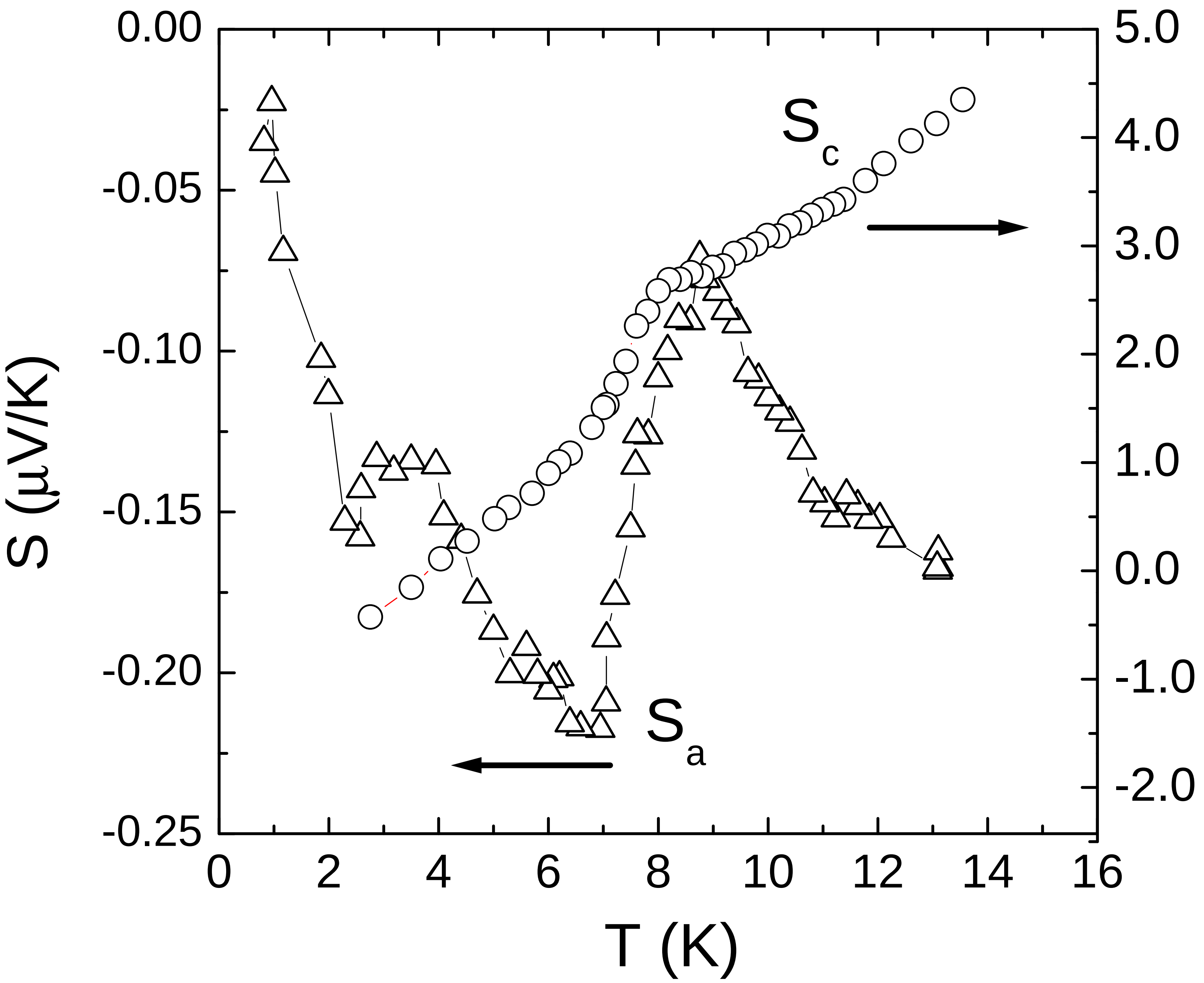

Choi et al., Fig. 3

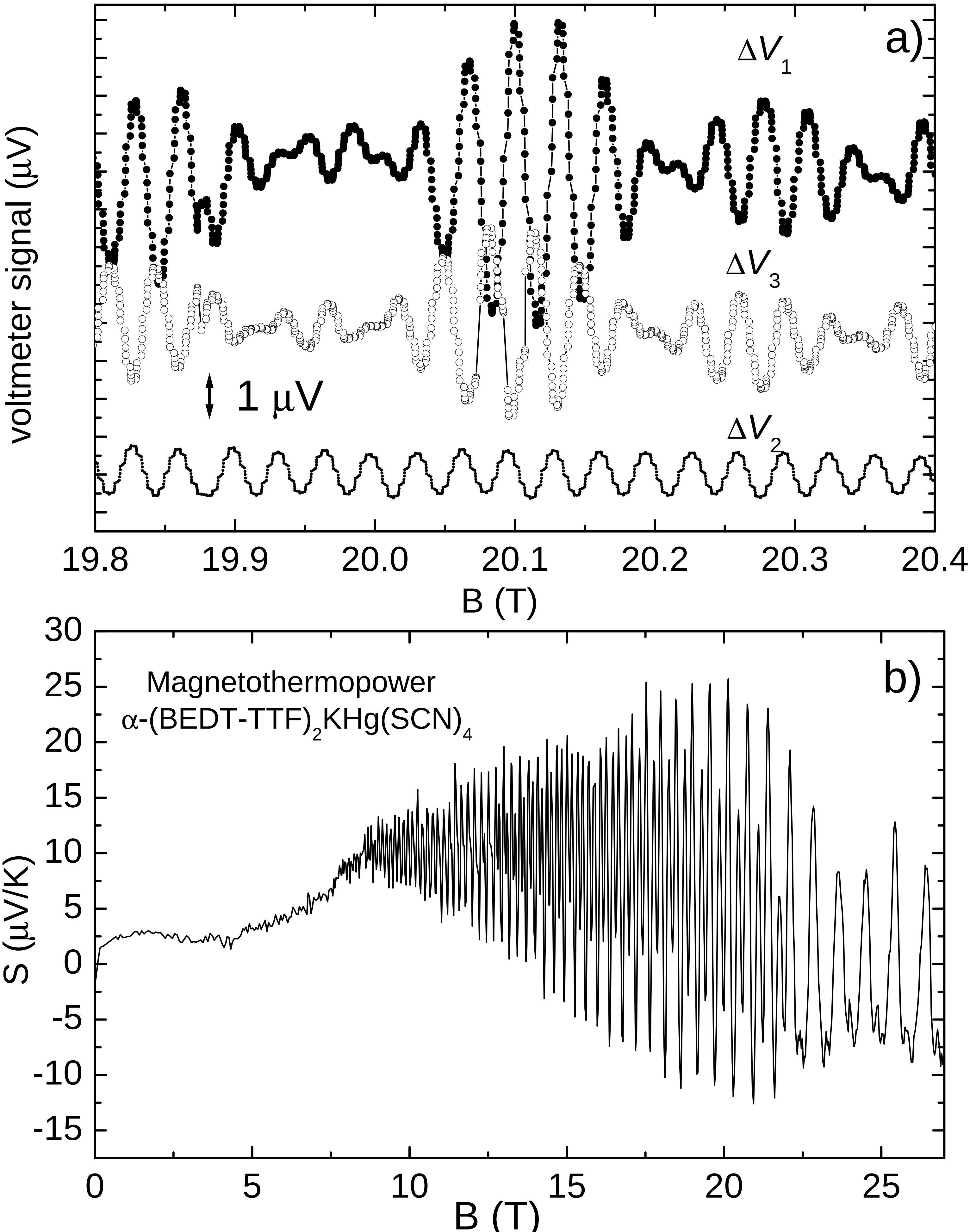

Choi et al., Fig. 4 a&b

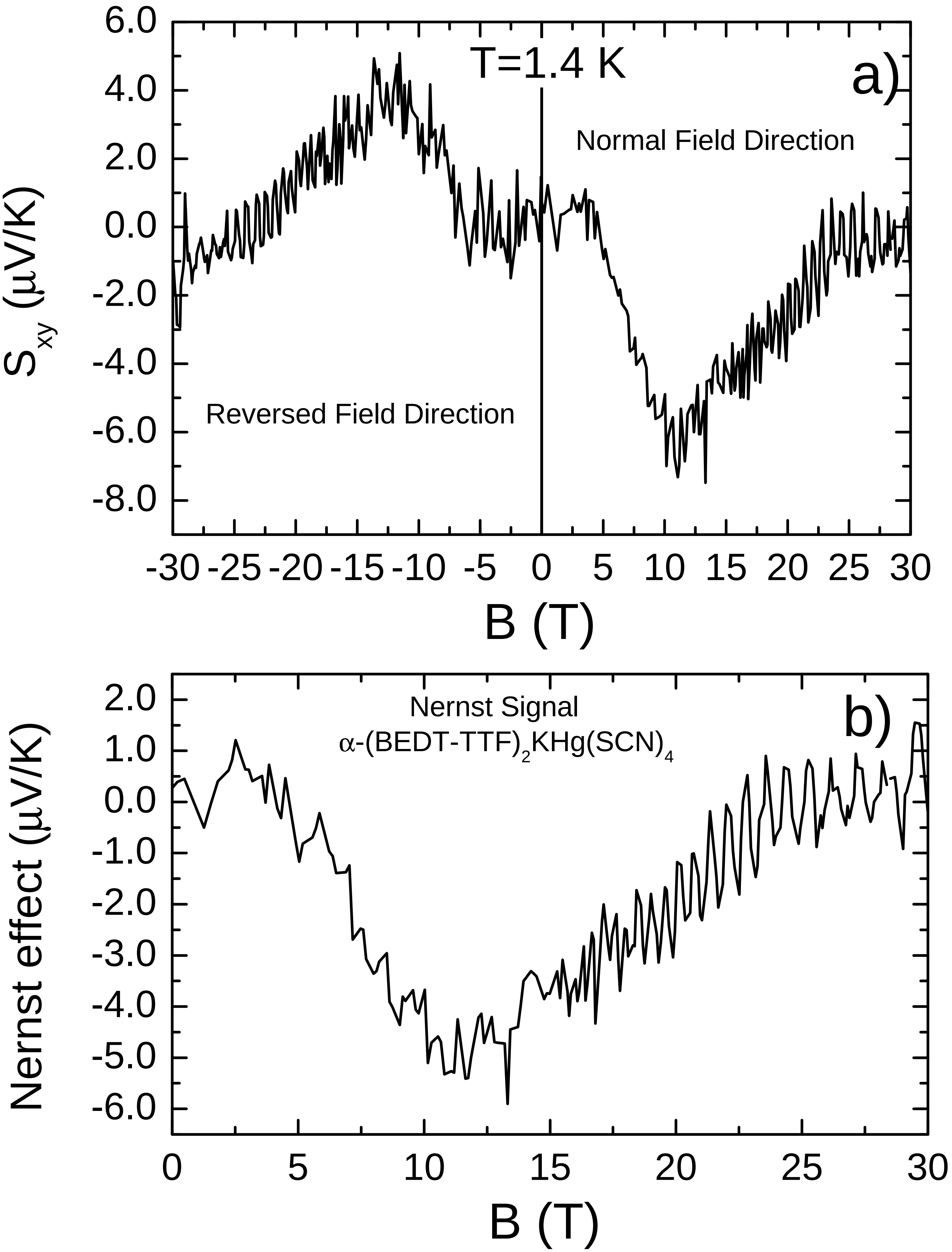

Choi et al., Fig. 5

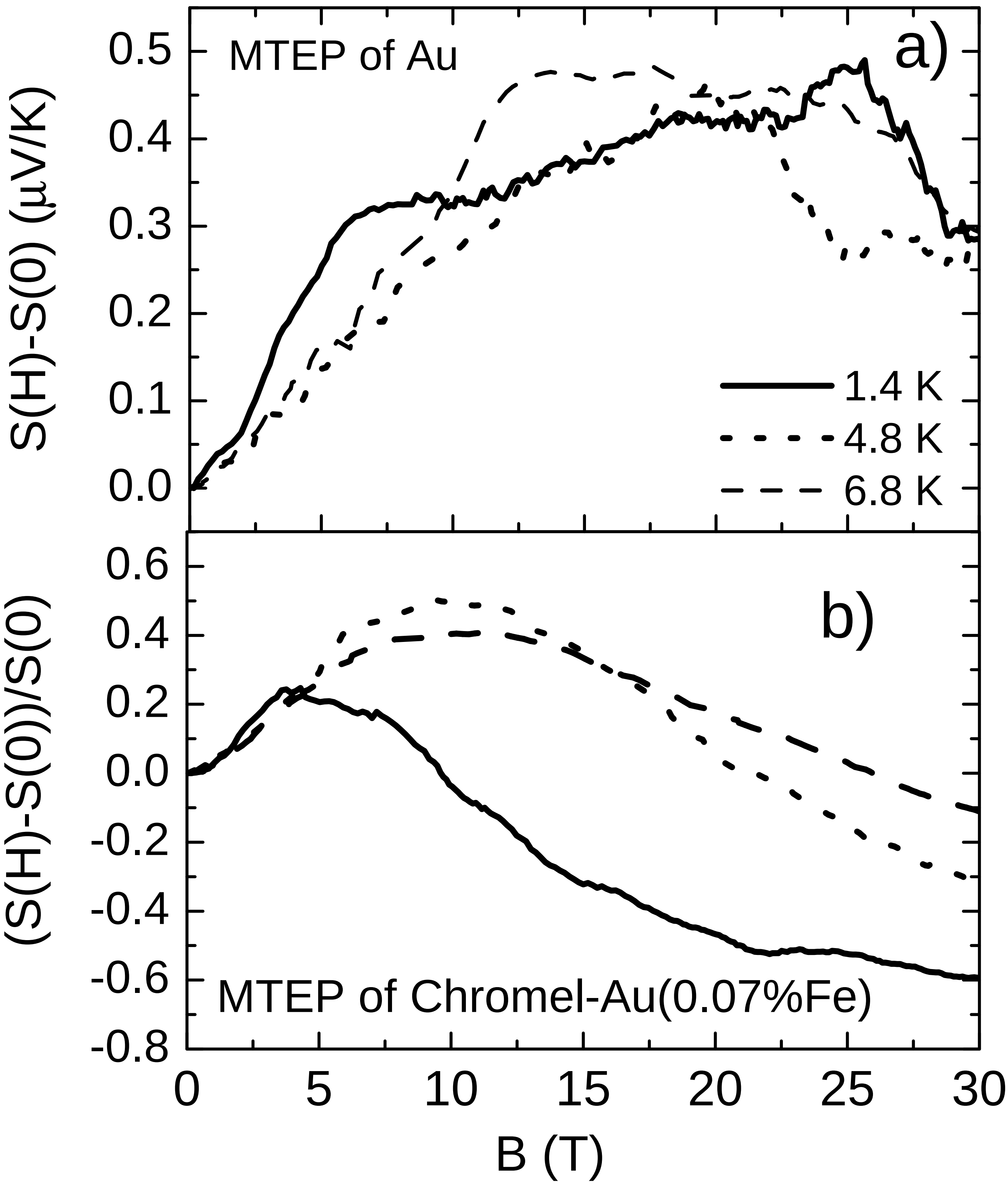

Choi et al., Fig. 6

# Low-frequency method for magnetothermopower and Nernst effect measurements on single crystal samples at low temperatures and high magnetic fields

E. S. Choi

*Research Institute for Basic Sciences, Ewha Womans University, Seoul 120-750, Korea and National High Magnetic Field Laboratory, Florida State University, Tallahassee, Florida 32310*

J. S. Brooks and J. S. Qualls*

*National High Magnetic Field Laboratory and Physics Department, Florida State University, Tallahassee, Florida 32310*

Y. S. Song

*Texas Center for Superconductivity, University of Houston, Houston, TX 77204-5932*

ABSRACT:

We describe an AC method for the measurement of the longitudinal ($S_{xx}$) and transverse ($S_{xy}$, i.e. Nernst) thermopower of mm-size single crystal samples at low temperatures (T<1 K) and high magnetic fields (B>30 T). A low-frequency (33 mHz) heating method is used to increase the resolution, and to determine the temperature gradient reliably in high magnetic fields. Samples are mounted between two thermal blocks which are heated by a sinusoidal frequency $f_0$ with a $\pi/2$ phase difference. The phase difference between two heater currents gives a temperature gradient at $2f_0$. The corresponding thermopower and Nernst effect signals are extracted by using a digital signal processing method due. An important component of the method involves a superconducting link, $YBa_2Cu_3O_{7+\delta}$ (YBCO) , which is mounted in parallel with sample to remove the background magnetothermopower of the lead wires. The method is demonstrated for the quasi two-dimensional organic conductor $\alpha$-$(BEDT\text{-}TTF)_2KHg(SCN)_4$, which exhibits a complex, magnetic field dependent ground state above 22.5 T at low temperatures.

Corresponding Author:
Prof. James Brooks, Physics
NHMFL/Physics
1800 E. Paul Dirac Dr.
Tallahassee FL 32310 USA
brooks@magnet.fsu.edu
Phone: 1-850-644-2836 (-5038 fax)

I. INTRODUCTION

The application of a thermal gradient ($\Delta T$) across a conducting material leads to a corresponding potential difference, or thermo-electric power (TEP, or thermopower). Thermopower measurements yield information about both thermodynamic properties and the transport properties of carriers. Advantages of TEP include the zero-current nature of the measurement, and its sensitivity to band structure, especially in the case of anisotropic (low dimensional Fermi surface) materials. Following Mott and Jones, the thermopower[1] of a metal may be expressed as

$$S = \frac{\pi^2 k_B^2 T}{3e}\left(\frac{d\ln n(E)}{dE} + \frac{d\ln v^2(E)}{dE} + \frac{d\ln \tau(E)}{dE}\right)_{E=E_F} \qquad (1)$$

where $n(E)$ is the density of the states, $v(E)$ is an average charge velocity, and $\tau(E)$ is the carrier scattering relaxation time. As we will show in the present application, the derivative of $n(E)$ at the Fermi energy will lead to large oscillations in systems where Landau quantization of the electronic energy levels occurs at high magnetic fields[2]. Hence oscillations in TEP associated with the de Haas van Alphen effect can be observed.

In a typical experimental setup, similar to that shown in Fig. 1, a sample is connected between two thermal platforms. A temperature difference is applied by heating one of the platforms, and $\Delta T$ is measured either between the platforms, or at points on the sample. The electric potential difference $\Delta V$ is measured with contact leads on the sample. In general, the apparatus is in weak thermal contact with a reference bath at a variable temperature $T$. As with other such measurements on small samples (e.g. specific heat), the effects of the addenda, and the magnetic field dependence of the sensors, cannot be negelected. Previous magneto-thermopower (hereafter MTEP) measurements at low temperatures and in high magnetic fields have addressed experimental issues such as the magnetothermopower of lead wires[3,4]. By using well-studied elemental metal wires of copper or gold, and high $T_c$ superconductors (where S = 0 for $T << T_c$ and $B << B_{c2}$), these background contributions may be sorted out. For long, thin samples (mm size samples with 10:1 to 100:1 aspect ratios), an AC technique has been used to measure MTEP for wide range of temperature[5]. But these techniques cannot be easily adapted to small single crystals with 1:1 aspect ratios, as in the case of the quasi-two dimensional "ET" organic conductors (see results section), where an accurate determination of $\Delta T$ in high magnetic fields becomes difficult. Resel et al. introduced a MTEP measurement technique up to 17 Tesla and down to 3K where chromel-constantan thermocouples are used as voltage and $\Delta T$ leads simultaneously[6]. Their technique (as does ours) includes the alternate heating method ("seesaw heating") to increase the measurement accuracy. However, there are limits in their methods in the case of small samples, since the large absolute TEP of chromel wire can introduce substantial background signal, and the application of a thermocouple junction directly to a small sample, can cause complications.

The technique to be introduced in this paper utilizes a stable, alternating heating method at very low frequency. Here the lead wires are in-situ calibrated using a $YBa_2Cu_3O_{7+\delta}$ high $T_c$ superconductor sample as a reference. When combined with digital signal processing methods, our procedure leads to enhanced resolution and accuracy for the MTEP of small samples, with direct application to high magnetic field measurements.

## II. EXPERIMENTAL TECHNIQUE

### A. Measurement setup

Fig. 1 shows the schematic diagram of the MTEP and the Nernst effect measurement holder in a top-view, where the magnetic field is applied normal to the plane of the figure. The apparatus is held in a 10 mm diameter cylindrical copper holder that is sealed with a copper cap (with a threaded grease seal). The copper holder can be maintained at any temperature T between 300 K and 0.5 K in a standard cryogenic, high-field dewar arrangement. The integrity of the seal is checked by a small jump in an applied temperature gradient of the apparatus when the encapsulated air in the holder condenses out below 80 K. Since $^3$He exchange gas is used to cool the holder, superfluid leaks do not present a problem. Samples are mounted between two quartz blocks (2.9 x 2.4 x 1.0 $mm^3$), A and B, with the ends attached using Apiezon N grease[7]. Electrical contacts to the samples are made by 12.5 μm gold wires using silver (or carbon) paste. The electrical connection between the lead wires and the external wires is kept in an isothermal condition by thermally anchoring them to the copper holder. Chromel-Au(Fe0.07%) thermocouple wires[7], used to measure $\Delta T$ between the quartz blocks, are attached to the quartz blocks using Stycast 2850 GT epoxy[8]. To minimize the experimental inaccuracy, which may result from the difference of temperature between the quartz block and the sample, the temperature gradient was produced by heating quartz blocks at a low frequency of order 33 mHz. Two chip resistor heaters (220 Ω RuO2 miniature surface-mount resistors) are attached by Stycast 2850 GT epoxy[8] to the edge of quartz blocks to enhance the homogeneous heat conduction.

Two sinusoidal currents are applied by Keithley 220 programmable DC current sources[9] with same frequency $f_0$ but with a $\pi/2$ phase difference. The relation between the two heater currents and $\Delta T$ for ideal system (perfect heat conduction from the heater to the quartz block) is the following:

$$I_1(t) = I_0 \sin(2\pi f_0 t)$$

$$I_2(t) = I_0 \sin(2\pi f_0 t + \pi/2) = I_0 \cos(2\pi f_0 t) \qquad (2)$$

$$\Delta T(t) \propto (I_1^2 - I_2^2)\, R / C_p = I_0^2 R / C_p \times \sin(2\pi(2 f_0)t)$$

where $I_1$, $I_2$ are heater currents, $R$ is heater resistance, $C_p$ is the heat capacity of the quartz block and t is the time. The validity of last expression in Eq. (2), i.e. the assumption of equal power for both heaters, can be checked by the Fourier analysis of the voltage signal of thermocouple wires. If the values of $R$ and $C_p$ are not identical, $\Delta T$ will oscillate with frequency of $2f_0$ and $f_0$ as follows;

$$\Delta T(t) \propto A\sin(2\pi f_0 t) - (A+\delta)\cos(2\pi f_0 t) = A\sin(2\pi(2 f_0)t) - \delta\cos(2\pi f_0 t) \qquad (3)$$

By the spectrum analysis of $\Delta T$ through the Fourier transform, two peaks will appear at $f=f_0$ and $f=2f_0$. One can estimate the contribution of the non-identical heat transfer ($\sim\delta/A$) from the ratio of amplitudes of peaks. For our holder, the value was found to be about 0.02. Even the contribution from the non-identical heat transfer is substantial, it can be dealt with by an appropriate analysis of the signal, as discussed below. For the present case, we consider only the dominant $2f_0$ contribution. Since $\Delta T$ oscillates with $2f_0$ frequency, corresponding TEP and the Nernst voltage will also oscillate with the same frequency. This second harmonic detection has an advantage in reducing electrical cross-talk which may arise from single harmonic generation in the heaters.

Figs. 2a and 2b show the applied heater currents and the corresponding thermoelectric potential $\Delta V_1$ and the thermocouple *emf* $\Delta V_2$ as a function of the time. ($\Delta V_1$ and $\Delta V_2$ refer to voltage leads 1 and 2 as shown in Fig. 1.) $\Delta V_2$ , the measured *emf* of the Chromel-Au(Fe0.07%) thermocouple, is related to $\Delta T$ by $\Delta V_2 = - \Delta T \times S_{Ch\text{-}AuFe} + V_{offset}$, where $S_{Ch\text{-}AuFe}$ is the TEP of Chromel-Au(Fe0.07%) thermocouples[7] and $V_{offset}$ is the offset voltage which lies in the range of 0.1 ~ 0.2 μV at low temperature. $\Delta V_1$ and $\Delta V_2$ were measured with Keithley 2182 nanovoltmeters[9]. There is a slight phase difference (~ 14 degrees or about 0.6 s delay) between $\Delta T$ and $\Delta V_2$ due to the non-ideal heat conduction. However, since the signals are digitally averaged over 2 ½ periods (see below) to obtain the amplitudes of $\Delta T$ and $\Delta V_2$, the phase difference does not enter into the final TEP value. When $\Delta T$ is small compared to the measurement temperature $T$, the absolute TEP ($S_{xx}$) and Nernst voltage ($S_{xy}$) can be expressed as :

$$
\begin{aligned}
S_{xx}(T,B) &= \frac{V_1(T,B)}{\Delta T(T,B)} \times P_{xx}(T,B) + S_{Au}(T,B) \\
&= \frac{V_1(T,B)}{V_2(T,B)/S_{Ch-AuFe}(T,B)} \times P_{xx}(T,B) + S_{Au}(T,B) \qquad (4)
\end{aligned}
$$

$$
S_{xy}(T,B) = \frac{V_1(T,B)}{V_2(T,B)/S_{Ch-AuFe}(T,B)} \times P_{xy}(T,B)
$$

where $S_{Au}$ is the absolute thermopower of Au and $P$ is either +1 or –1 depending on the phase difference between $V_1(V_3)$ and $V_2$. $S_{Au}$ can be determined from the measurement of another sample (YBCO in this paper) whose value is known at a certain temperature and magnetic field. For the Nernst voltage, it is assumed that sample alignment is ideal so that there is no contribution from $S_{xx}$. When the misalignment is substantial, $S_{xy}$ can be obtained from the difference in the Nernst voltage for two magnetic field sweeps with opposite polarities.

The amplitude of oscillation ($V_1$ and $V_2$ in Fig. 2) can be determined from the discrete values ($\Delta V_i$) by the following formula[10] :

$$V = \sqrt{\sigma_x^2[H\{\Delta V_i - 2\mu_x(H(\Delta V_j))\}]\times 8/3}\times\sqrt{2} \qquad \text{(5)-1}$$

where $\sigma_x^2$ is the variance, $\mu_x$ is the mean value and $H(x_i)$ is the Hanning window defined by

$$H(x_i) = 0.5\times\{1-\cos(\frac{2\pi i}{\#\ of\ \ the\ elements})\} \text{ for } i = 0, 1, 2, \ldots, \text{n-1}. \qquad \text{(5)-2}$$

The Hanning window was used to separate AC signal from DC signal ($V_{offset}$), and hence the signal was compensated for the windowing effect by a multiplicative constant (8/3 in this case). Finally the root mean square value was used to extract an amplitude of the AC signal. Because this method does not discriminate oscillations with different frequencies (for example, dominant $2f_0$ signal and $f_0$ signal from non-identical heat transfer), this method has an advantage that the non-identical heat transfer term can be also considered. However, if there is a substantial low frequency noise, one should use digital bandpass filters or FFT analysis to obtain the amplitudes.

We note that there have been previous AC TEP measurements using the $2f_0$ mode by Kettler et al.[11], where limitations due to the heat capacity of the material and the characteristic times of thermal relaxation were identified. To overcome these problems, the excitation frequency $2f_0$ should be as small as possible. In our measurement setup, $2f_0$ was chosen to be 67 mHz, i. e., the oscillation period of the heater currents is 30 seconds, and the corresponding oscillation period of $\Delta T$ is 15 seconds. Our method for determining the frequency range where the AC and ideal DC methods coincide is shown in Fig. 2c. A suitable excitation frequency range for the apparatus in Fig. 1 is for $2f_0$ below 100 mHz. Above 100 mHz, the AC TEP increases as a function of frequency due to various thermal relaxation rates which are characteristic of the apparatus.

B. Measurements in magnetic field

The difficulties for the MTEP measurement comes from the field dependence of $S_{Au}(B)$ and $S_{Ch\text{-}AuFe}(B)$. To avoid the problem of $S_{Ch\text{-}AuFe}(B)$, we exploit the high reproducibility of $\Delta T$ for corresponding constant amplitudes of the heater currents. $V_2$ changes very little with time during a measurement for fixed temperature (typically less than 1 mK over a 20 minute period of measurement). Correspondingly, the change of $\Delta T$ is also very small for magnetic field sweeps , with less than 1% deviation in $\Delta T$ as compared with the zero field value . The change of $\Delta T$ will come from the field dependence of specific heat of the quartz block and the magnetoresistance of the heater resistor. The former is negligible for the quartz block and the latter can be calibrated at each magnetic field. Although the magnetoresistance ($MR = R(B)/R(B{=}0)$) of the chip resistor used in this measurement is very small ( for 30 T : ~ -1.5% at 4K, ~ -1.4% at 1.1K, ~ -2.0 % at 0.7 K), we also included the $MR$ effect in the determination of $\Delta T\ (T, B)$. Once $\Delta T(T, B)$ is determined, $S_{Au}(B)$ can be easily measured using a YBCO sample as a reference below its critical field, where $S_{YBCO}(B) = 0$. Therefore the MTEP and Nernst voltage can be written as

$$S_{xx}(T,B) = \frac{V_1(T,B)}{\frac{V_2(T,B=0)}{S_{Ch-AuFe}(T,B=0)} \times MR(T,B)} \times P_{xx}(T,B) + S_{Au}(T,B)$$

$$S_{xy}(T,B) = \frac{V_3(T,B)}{\frac{V_2(T,B=0)}{S_{Ch-AuFe}(T,B=0)} \times MR(T,B)} \times P_{xy}(T,B) \qquad (6)$$

where

$$S_{Au}(T,B) = 0 - \frac{V_1(T,B)}{\frac{V_2(T,B=0)}{S_{Ch-AuFe}(T,B=0)} \times MR(T,B)} \times P_{xx-YBCO}(T,B) \qquad (7)$$

for $T < T_c$ and $B < B_c$ of YBCO.

## III. EXPERIMENTAL RESULTS

To demonstrate the techniques described here, we consider the MTEP and the Nernst effect of organic conductor α-(BEDT-TTF)$_2$KHg(SCN)$_4$ . The material α-(BEDT-TTF)$_2$KHg(SCN)$_4$ is a well known quasi two-dimensional organic conductor[12] which shows metal - density wave transition around $T$ = 8 K, and which has a magnetoresistance anomaly at 22.5 T (below 8 K) where there is a magnetic field induced change in the electronic structure. The typical size of the sample is about 1.4×0.6×0.26 mm$^3$ , with a plate-like morphology. The sample is mounted so that the temperature gradient is in the plane of the conducting layers (a or c-axis), with the field perpendicular to the conducting layers (b-axis). A polycrystalline YBCO sample, of comparable dimensions, was used for in-situ calibration. The sample holder was attached to the probe of $^3$He cryostat and a 30

T resistive magnet at the National High Magnetic Field Laboratory was used for the high field measurements.

Fig. 3 shows zero field TEP results for the crystallographic *a*- and *c*-axis, i.e. $\Delta T$ is parallel with *a*- and *c*-axis respectively. As mentioned in the introduction, the TEP is sensitive to the anisotropy of the band structure (and therefore the Fermi surface), hence it depends on the direction of $\Delta T$ with respect to the crystallographic axes. Structure in the TEP, due to the opening of a partial gap, is clearly seen around 8K for both axes. The sum of the amount of jump is about 1 μV/K, which is in reasonable agreement with the heat capacity measurement results[13].

For MTEP and the Nernst effect measurement, the magnetic field is swept very slowly (0.042 T/min) for fixed temperatures. Fig 4 (a) shows the raw data of $\Delta V_1$, $\Delta V_2$ and $\Delta V_3$ for T = 0.7 K. When the heater power is about 220 μW, $\Delta T$ is 0.085K at zero field and it is assumed to decrease to 0.083 K at B = 27 T due to the negative MR of the chip resistor. The oscillation of $V_1$ and $V_3$ is huge and the change of polarity ($P(T, B)$ of Eq. (4) and(6))can be also seen from the raw data.. The derived MTEP data are shown in Fig. 4 (b). Each MTEP data point is obtained using Eq. (5) - (7) by averaging over ~ 100 raw data points.

Fig. 5 shows the Nernst voltage ($S_{xy}$) for positive and negative magnetic field sweep and the corresponding Nernst effect at T = 1.4K. For this case, the xy sample electrode alignment is quite good and quantum oscillations can be clearly seen, even though the final results are obtained by subtracting $S_{xy}$ ($B$<0) from $S_{xy}$ ($B$>0), and dividing by 2. The corresponding period of quantum oscillations can be obtained from the Fourier transform, and it is found to be ~ 670 T, which agrees with the value measured from the magnetoresistance and magnetization experiments[14].

Finally, we describe the MTEP of the thermocouple (Chromel-AuFe) and electrical (Au) leads used in the present study. This is done under the assumption that $\Delta T$ depends only on the heater power (taking into account the negative MR), and assuming that the magnetic field effect on the heat capacity of quartz and stycast epoxy is negligible. Then, for a fixed heater power ($\Delta T$) and temperature T, $\Delta V_4$ on the YBCO sample and $\Delta V_2$ on the quartz blocks are measured as a function of magnetic field. $S_{Au}(T, B)$ can then be determined directly, and from Eq. (7), we can determine $S_{Ch\text{-}AuFe}(T, B)$. Fig. 6 shows the results at different temperatures as a function of the magnetic field. For the MTEP of $S_{Ch\text{-}AuFe}$, the overall behavior is quite similar to the previous determinations[15] in which the MTEP was measured by a different method up to $B$ = 14 T.

ACKNOWLEDGEMENTS

This work was supported in part by NSF-DMR95-10427 and DMR99-71474. The work was carried out at the National High Magnetic Field Laboratory, supported by a contractual agreement between the State of Florida and the NSF through NSF-DMR-95-27035. E. S. Choi was financially supported by the KOSEF postdoctoral fellowship program.

REFERENCES

* Present address: Physics Dept., Wake Forest University, Winston-Salem, NC 27109.

Figure captions

Fig. 1. Diagram of the measurement holder (the outer diameter of the cylindrical copper holder is 10 mm). A : Cu heat sink, B : quartz blocks, C : heaters. 1 : thermopower leads of sample, 2 : Chromel-Au(Fe0.07%) thermocouples for $\Delta T$ leads, 3: Nernst voltage leads of sample, 4 : thermopower leads of reference YBCO sample.

Fig. 2. a) Heater currents and b) $\Delta V_1(\Delta V_2)$ as a function of the time. The period of the heating cycle is 30 seconds and the corresponding periods of oscillation of temperature gradient and thermopower signal are 15 seconds. c) $S_{AC}/S_{DC}$ vs. frequency method used to determine the optimum frequency range where $S_{AC}/S_{DC} \approx 1$ for the TEP measurements.

Fig. 3. Zero-field thermopower results of $\alpha$-(BEDT-TTF)$_2$KHg(SCN)$_4$ as a function of temperature. A gap opens in the quasi-one dimensional part of the Fermi surface near 8 K, which is seen as a peak in the a-axis data. Filled circles : $\Delta T \parallel a$-axis, open circles : $\Delta T \parallel c$-axis.

Fig. 4. Magnetothermopower. (a) $\Delta V_1$, $\Delta V_2$ and $\Delta V_3$ curves under magnetic field for $\alpha$-(BEDT-TTF)$_2$KHg(SCN)$_4$ at T = 0.7 K. (b) Derived magnetothermopower results. Note the narrow range of field in (a) , which corresponds to only a few quantum oscillations in (b).

Fig. 5. Nernst effect. a) $S_{xy}$ signal of $\alpha$-(BEDT-TTF)$_2$KHg(SCN)$_4$ for normal and reversed field sweeps at T=1.4 K. The large asymmetry, even in the raw data, indicates a significant xy component of the thermopower. b) Corresponding Nernst voltage (= $(S_{xy}(B>0)-S_{xy}(B<0))/2$).

Fig. 6. a) ($S(B)-S(B=0)$) of Au and b) ($S(B)-S(B=0))/S(B=0)$ of Chromel-Au(Fe0.07%) thermocouples.